\documentclass[11pt,a4paper]{article}
\pdfoutput=1
\usepackage{jheppub}
\usepackage{graphicx}
\usepackage{verbatim}
\usepackage{wasysym}
\newcommand{\ba}{\begin{eqnarray}}
\newcommand{\ea}{\end{eqnarray}}
\newcommand{\no}{\nonumber}
\newcommand{\be}{\begin{equation}}
\newcommand{\ee}{\end{equation}}
\newcommand{\bea}{\begin{eqnarray}}
\newcommand{\eea}{\end{eqnarray}}

\title{A Heavy ``Neutralino" in Warped Extra dimensions
}
\date{\today
}
\author{
Luca Vecchi}
\affiliation{Maryland Center for Fundamental Physics,\\ Department of Physics, University of Maryland\\
College Park, MD 20742, USA}
\emailAdd{vecchi@umd.edu}

\abstract{Generic extensions of the Standard Model that respect baryon and lepton numbers have accidentally stable particles. Typical examples are the lightest exotic neutral fermion, or ``neutralino", and fields with non-trivial lepton and baryon charges. In this paper we identify the accidentally stable neutralino with the dark matter, and discuss its phenomenology in the framework of warped extra dimensions. We find that annihilation into other Kaluza-Klein resonances is often allowed and very efficient. The observed dark matter abundance may then be obtained with couplings of order unity and a compactification scale above the TeV. Light dark matter is also possible in the presence of unsuppressed couplings to the Higgs boson. In this latter case dark matter direct detection experiments will soon be able to probe a significant portion of the parameter space. This analysis suggests that dark matter is a natural feature of realistic models with warped extra dimensions. 
}
\begin{document}
\maketitle
%%%%%%%%%%%%%%%%%%%%%%%%%%%%%%%%%%%%%%%%%%%%%%%%%%%%%%%%%%%%%%%%%%%
%
%%%%%%%%%%%%%%%%%%%%%%%%%%

\section{Baryon and Lepton numbers imply Dark Matter}

There are several phenomenological and theoretical reasons to expect a new physics threshold well below the Planck scale. Yet, this expectation faces a serious challenge when compared to experimental evidence. At low energies, an arbitrary extension of the Standard Model (SM) characterized by a mass scale $\Lambda$ will generate, among others, operators such as
\ba\label{BL}
c_\nu\frac{\ell\ell HH}{\Lambda},~~~~~~~~c_{p}\frac{qqq\ell}{\Lambda^2},
\ea
with $q$ and $\ell$ quark and lepton fields and $H$ the Higgs doublet. The first interaction gives a Majorana mass to the SM neutrinos. Requiring this contribution does not exceed the largest observed neutrino mass, of order $(\Delta m^2_{\rm atm})^{1/2}=0.05$ eV, gives a lower bound $\Lambda\geq 6|c_\nu|\times10^{14}$ GeV. The second mediates proton decay and is  very strongly constrained as well. Current bounds on the proton lifetime roughly require $\Lambda\gtrsim\sqrt{|c_p|}10^{16}$ GeV. 

These remarkable experimental achievements suggest that either the new physics lives at currently unaccessible energies, or $\Lambda$ is relatively low but $|c_\nu|$ and $|c_p|$ are much smaller than unity. The latter possibility is certainly more exciting, and is naturally realized if the new dynamics approximately respects the SM lepton number, or alternatively both $U(1)_B$ and $U(1)_L$. 

Strictly speaking, baryon number is not necessary to suppress~(\ref{BL}), and moreover we know that it was not a good symmetry in the early Universe since $U(1)_B$ violation is actually one of the necessary conditions for baryogenesis. However, allowing large violations of $U(1)_B$ is not always possible. If for example a primordial $B-L$ asymmetry was generated at some high scale, say via leptogenesis, new sources of $U(1)_B$ (or $U(1)_L$) violation in thermal equilibrium above the electroweak phase transition would completely wash it out. In this case one would be forced to consider scenarios of low scale baryogenesis or cosmologies with a low reheat temperature.~\footnote{Collider bounds on $pp\to K^+K^+$ and $n-\overline{n}$ oscillation are satisfied by a relatively low scale $\Lambda\gtrsim1$ TeV, and do not seem as relevant as the one mentioned above.}

The alternative is to make no assumption regarding the scale of baryogenesis and inflation, and instead require that the new physics be (approximately) invariant under {\emph{both}} lepton and baryon numbers. This symmetry may result from an ``accident" as in the SM, or might be the indication of an underlying more fundamental symmetry. Baryogenesis in this case can safely occur at high temperatures $\gtrsim10^{16}$ GeV, without being wiped out by low energy effects. Yet, this scenario is also compatible with low scale baryogenesis, if (not too) small $U(1)_B$-violating couplings turn on after the electroweak sphalerons have decoupled.

There is a priori no reason to prefer models in which $U(1)_B\times U(1)_L$ is a (approximate) symmetry compared to scenarios with lepton number and low energy baryogenesis. However, we know from our experience in model building that the former class of theories has often an attractive spin-off: generic extensions of the SM satisfying $U(1)_B\times U(1)_L$ naturally include stable particles, and hence dark matter (DM) candidates.

The most notable example is of course realized in the minimal supersymmetric SM, where a ${\mathbb Z}_2$ symmetry, R-parity, implies baryon and lepton numbers at the renormalizable level. This relation is not just a feature of the MSSM, though. In any theory satisfying $U(1)_B\times U(1)_L$, the lightest exotic neutral fermion will automatically be stabilized by an accidental ${\mathbb Z}_2$. This is seen by observing that conservation of angular momentum implies that any decay mode of that particle must have an odd number of SM fermions in the final state, and that such a configuration necessarily carries baryon or lepton charges. Without loss of generality, we can prove the existence of an accidental ${\mathbb Z}_2$ by writing down the most general electric and color neutral operator involving protons $p$, neutrons $n$, electrons $e$, neutrinos $\nu$, and our exotic fermion $X$ (we do not include photons and derivatives to simplify our notation, whereas heavy leptons and other hadrons are unstable and can be thought of as an array of $e,\nu,p,n$). Imposing Lorentz and $U(1)_B\times U(1)_L$ invariance this reads
\ba
(p\bar ne \bar\nu)^mX^{2n}\times(\dots),
\ea
with $m,n$ integers and the dots standing for terms with arbitrary powers of proton anti-proton pairs, neutron anti-neutron pairs, etc., that are trivially neutral. The fact that $X$ appears in even powers is sufficient to show the existence of an accidental DM parity symmetry.

We can further generalize this result in scenarios where $U(1)_B\times U(1)_L$ is embedded into a larger symmetry or a local invariance. Typical realizations are weakly coupled theories with {\emph{gauged}} $U(1)_B\times U(1)_L$~\cite{B&L}. Models with warped extra dimensions belong to the same class. In fact, quantum gravity effects are expected to violate continuous global charges, so $U(1)_B\times U(1)_L$ invariance can only be implemented as a {\emph{local}} symmetry in a gravitational theory. A realistic Randall-Sundrum scenario then requires the symmetry be spontaneously broken, say on the UV brane (see for instance~\cite{Agashe:2002pr} and~\cite{Agashe:2004ci} for a discussion of gauged $U(1)_B$ on warped backgrounds).

In all these scenarios it makes sense to talk of particles $X$ with definite charges $(q_B,q_L)$ under $U(1)_B\times U(1)_L$. Now, assuming that $X$ is an electric and color neutral fermion, and imposing Lorentz as well as $U(1)_B\times U(1)_L$ invariance, we find that the decays
\ba
X\to (pe)^{m}n^{n}\nu^{p}\dots
\ea
are forbidden whenever~\footnote{A similar argument was presented in~\cite{Graesser:2011vj} for models with local $U(1)_B$.}
\ba\label{stability1}
q_B+q_L\neq{\rm odd}~~~~~&{\rm or}&~~~~~~q_B-q_L\neq{\rm odd}.
\ea 
Note that the neutralino --- with $(q_B,q_L)=(0,0)$ --- is just a trivial example satisfying condition~(\ref{stability1}). Repeating the same exercise for $X$ a color and electric neutral boson we find that $X$ is stable when
\ba\label{stability2}
q_B+q_L\neq{\rm even}~~~~~&{\rm or}&~~~~~~q_B-q_L\neq{\rm even}.
\ea
Stability of these DM candidates is again guaranteed by an {\emph{accidental}} DM parity symmetry. One can prove this statement along the lines outlined for the neutral fermion case. 

Note that our perspective is reversed compared to what usually assumed in the model-building literature. In our view the baryon and lepton numbers are ``fundamental" symmetries of Nature whereas the DM parity is ``accidental". While remaining completely agnostic regarding the UV origin of $U(1)_B\times U(1)_L$, we found that the very existence of such a phenomenologically motivated symmetry is sufficient to imply a DM parity. The opposite is known not to hold in general. In the MSSM, for example, R-parity ensures the existence of a stable DM candidate, but does not forbid dangerous dimension-5 operators mediating proton decay. In the present framework, it is the accidental DM symmetry that only holds as long as $U(1)_B\times U(1)_L$ is exact. Once a breaking of baryon and lepton numbers is introduced (for example to achieve a successful baryogenesis), one has to make sure that the DM remains long lived on cosmological time scales. This can be a non-trivial constraint, as we will see.

Among the most motivated low energy extensions of the SM are Supersymmetry and models with warped extra dimensions (that are equivalent to models with TeV scale compositeness). DM is a welcome feature of the $U(1)_B\times U(1)_L$ invariant MSSM. However, minimal warped extra-dimensions do not have an obvious DM candidate. Yet, according to our argument, any realistic $U(1)_B\times U(1)_L$ invariant extension generically will. 

There are many reasons to believe that minimality of the field content is not a justifiable guiding principle in Randall-Sundrum scenarios. Perhaps the most obvious is that these scenarios are inherently low energy effective field theories. The are other reasons that more directly relate with DM, though. For example, unification and baryogenesis are not realized in minimal scenarios, and new ingredients must be added. As shown in~\cite{Agashe:2004ci}, and in agreement with our conclusions, this can naturally lead to the appearance of stable particles in models with a long lived proton. Another motivation to go beyond the minimal scenarios considered so far in the literature is that new fields charged under the SM are actually required to cancel the (brane-localized) $U(1)_B\times U(1)_L$ anomalies. Some of these fields are expected to satisfy conditions~(\ref{stability1}) or (\ref{stability2}).

Rather than attempting to motivate a specific warped 5D scenario, it would be useful to first address the viability of the DM candidates resulting from~(\ref{stability1}) (\ref{stability2}) in a model-independent way. Following this logic, one may simply introduce a DM candidate ``by hand" and investigate under which conditions this field can account for the observed DM abundance. In this paper we begin this study by discussing the phenomenology of the neutral fermion candidate, i.e. a fermion $X$ with charges $(0,0)$ that is accidentally stable according to~(\ref{stability1}). The physics of DM with non-trivial charges under $U(1)_B\times U(1)_L$ is very different, and will be presented elsewhere.

We assume $X$ is the lightest Kaluza-Klein (KK) mode of a bulk field. Because the 5D field is neutral we expect both Dirac and Majorana 5D masses. We present an analysis of the KK reduction, as well as estimates for the DM mass, in section~\ref{sec:AdS}. In the appendix we extend the holographic interpretation of~\cite{Henningson:1998cd}\cite{Mueck:1998iz}\cite{Contino:2004vy} to 5D fermions with non-vanishing bulk Majorana mass. 

The central part of the paper is summarized in sections~\ref{sec:relic} and~\ref{sec:signatures}. There, we discuss the DM relic density, and the main experimental signatures of the ``heavy neutralino". We will see that the simple framework considered here, together with the effective field theory principle that all couplings allowed by the symmetries are present, suffice to obtain a realistic model for DM. This motivates our claim that DM is a generic feature of warped extra dimensions.

\section{A ``Neutralino" on AdS$_5$}
\label{sec:AdS}

Fermionic DM in warped extra dimensions have been studied before. Ref.~\cite{Agashe:2004ci} is the only previous work where a clear connection between proton stability and DM is made in this context. That paper discusses grand unification in Randall-Sundrum scenarios, and the importance of approximate baryon and lepton numbers is emphasized. The authors postulate a ${\mathbb{Z}}_3$ symmetry to forbid proton decay, and find a DM candidate with $U(1)_B\times U(1)_L$ charges $(1/3,0)$.~\footnote{We see in~(\ref{stability1}) that this particle would in fact be accidentally stable in a world with exact baryon and lepton numbers.} The phenomenology of our DM is however very different from that in ref.~\cite{Agashe:2004ci}. This can be traced back to the fact that in the present paper $X$ is a Majorana field while the DM was a Dirac particle in~\cite{Agashe:2004ci} (as required by ${\mathbb{Z}}_3$ invariance).

In~\cite{Kadota:2007mv} a light sterile Majorana neutrino was proposed as warm/cold DM. Its couplings to the bulk fields, with the exception of the gravitational interactions, were assumed to be negligible, so its relic density was determined by the late decay of bulk gravity. The DM is truly a SUSY neutralino in~\cite{Knochel:2008ks}. Finally, the DM candidate of~\cite{Carena:2009yt} was the lightest particle carrying a ``fundamental" ${\mathbb Z}_2$ symmetry.

Our approach is qualitatively different from these works in that no new symmetry is assumed beyond (approximate) $U(1)_B\times U(1)_L$ invariance. The results presented here are also quantitatively different because we find that the natural mass range for these DM candidates is multi-TeV.

\subsection{The DM mass}
\label{sec:spectrum}

The most general action for a neutral 5D fermion $\Psi$ contains both a Dirac and a Majorana mass terms:
\ba
-M_D\overline{\Psi}\Psi-\left(\frac{M_M}{2}\overline{\Psi}\Psi^c+{\rm hc}\right).
\ea
Here we defined $\Psi^c\equiv C_5\Psi^*$, with $C_5$ the 5D complex conjugation matrix. An explicit representation is $C_5=-i\gamma^5\gamma^2$ (see Appendix~\ref{app:spectrum}). Note that by a field redefinition the 5D Majorana mass $M_M$ can be made real. We will assume $M_M^*=M_M$ in this section (for a more general analysis we refer the reader to the appendix).

To perform the KK reduction it is convenient to introduce the following notation:
\ba\label{expl}
\Psi=
\left( \begin{array}{c}  \chi \\
\epsilon\psi^*
\end{array}\right),~~~~~~~~~~~\Psi^c\equiv -i\gamma^5\gamma^2\Psi^*=\left( \begin{array}{cc}  \quad & \epsilon \\
\epsilon & \quad
\end{array}\right)\Psi^*=
\left( \begin{array}{c}  -\psi \\
\epsilon\chi^*
\end{array}\right),
\ea
where $\chi,\psi$ are 4D left-handed spinors and $\epsilon=i\sigma^2$. With this notation, and specializing on a 5D background with conformal metric 
\ba
ds^2=a^2(z)(dx^\mu dx_\mu-dz^2),~~~~~~~~~~a(z)=\frac{L}{z},~~~~~~~~~z_{\rm UV}\leq z\leq z_{\rm IR},
\ea
the most general 5D action quadratic in $\Psi$ can be written as (see Appendix~\ref{app:spectrum})
\ba\label{5DMajo}
{\cal L}_{5D}
&=&a^4\left[i\chi^\dagger\bar{\sigma}^\mu\partial_\mu\chi+i\psi^\dagger\bar{\sigma}^\mu\partial_\mu\psi+\frac{1}{2}\left(\psi^t\epsilon\partial_z\chi-\partial_z\psi^t\epsilon\chi+{\rm hc}\right)\right]\\\no
&+&a^5\left[\frac{c_D}{L}\psi^t\epsilon\chi+\frac{c_M}{2L}\left(\chi^t\epsilon\chi-\psi^t\epsilon\psi\right)+{\rm hc}\right],
\ea
where a 4D total derivative has been dropped, and
\ba
c_{D,M}\equiv M_{D,M}L. 
\ea
In addition to~(\ref{5DMajo}) there could be boundary terms, which can in general relate the left and right handed components of $\Psi$. We will consider the class of boundary conditions $\psi\propto\chi$ in the following.

In complete generality the 4D spectrum can be assembled into a Kaluza-Klein tower of Majorana particles. The lightest mode will be our DM candidate. We denote it by the 4-component Majorana fermion $X=-i\gamma^2X^*$.

For real $M_M$ the eigenvalue problem can be written as (see Appendix~\ref{app:spectrum} for the generalization to complex $M_M$)
\ba\label{EOMc}
\Delta\left( \begin{array}{c}  
\chi_n\\
\psi_n
\end{array}\right)=m_n
\left( \begin{array}{c}  
\chi_n\\
\psi_n
\end{array}\right),
\ea
with~\footnote{The Pauli matrices are here called $\tau^a$ to distinguish them from $\epsilon=i\sigma^2$, which acts on the spinor indices.}
\ba\label{Delta}
\Delta\equiv i\tau^2\left(\partial_z+2\frac{\partial_za}{a}\right)-a\left(\tau^1M_D+\tau^3M_M\right),
\ea
plus the appropriate boundary conditions. Eq.~(\ref{EOMc}) generalizes the equations of motion for a fermion in AdS$_5$ first studied in~\cite{Grossman:1999ra}\cite{Gherghetta:2000qt}. 

The case $M_M\neq0$ was previously studied in~\cite{Huber:2003sf}, however our equations do not agree (they would if the 5D charge conjugation matrix $C_5$ is replaced by the 4D version $C_4=-i\gamma^2$). The mass spectrum and the wavefunctions that we derive in the present paper are therefore different from those of~\cite{Huber:2003sf} in the presence of a bulk mass $M_M\neq0$. For example, the authors of~\cite{Huber:2003sf} find that the zero mode solutions are oscillating functions of $M_M$, while eq.~(\ref{EOMc}) implies a power law behavior for any complex $M_M$ (see eq.~(\ref{zero}) below). We suspect the origin of this disagreement is in the definition of the complex conjugation matrix employed in~\cite{Huber:2003sf}. Our definition is consistent with the absence of Majorana fermions in 5D (see also eq.~(\ref{expl})).

An equivalent expression for~(\ref{EOMc}) can be obtained by writing the equations of motion in terms of $\Sigma=\psi/\chi$ and $\chi$ (or similarly $\Sigma$ and $\psi$), see eq.~(\ref{eqSigma}). This will come handy when discussing some properties of the massive spectrum (see below) and the holographic interpretation (Appendix~\ref{app:AdS/CFT}). However, $\Sigma$ is not very useful when it comes to solving the system numerically, since it contains physical poles.

The spectrum has both positive and negative eigenvalues $m_n$. In the absence of a Majorana mass the spectrum consists of Dirac pairs of opposite chirality. Thus, for any positive $m_n$, the solution consistently reveals a corresponding eigenvalue $-m_n$, characterizing the (equal) mass of the opposite chirality state. This may be understood mathematically by observing that the system satisfies the symmetry 
\ba\label{symmetry}
(\frac{\psi_n}{\chi_n},m_n,c_D,c_M)\to(-\frac{\psi_n}{\chi_n},-m_n,c_D,-c_M),
\ea
where $c_{D,M}=M_{D,M}L$. As soon as $c_M\neq0$ the degeneracy is continuously lifted and the 4D spectrum becomes a tower of Majorana particles. Thanks to~(\ref{symmetry}), the squared masses are symmetric under $c_M\to-c_M$, so that it is sufficient to analyze the positive branch $c_M>0$. The reason is that the sign of $c_M$ is unphysical, and can be arbitrarily chosen.

Explicit solutions for the massless modes are obtained straightforwardly from the equations of motion:
\ba\label{zero}
\chi_0=Az^{2-c}+\frac{c_M}{c+c_D}B z^{2+c}~~~~~~~~\psi_0=-\frac{c_M}{c+c_D}Az^{2-c}+B z^{2+c},
\ea
where we defined $c=\sqrt{c_D^2+c_M^2}$. Depending on the boundary conditions, this mode may or may not propagate. We will discuss this below.

For the massive spectrum there exists no analytic solution of~(\ref{EOMc}) when $c_{D,M}$ are both non-vanishing (there are when $c_D=0$, but these are not very relevant physically because there is no symmetry naturally leading to that limit). Yet, one can still infer some qualitative features by just staring at the equations. For example, the massive spectrum in the asymptotic regime $m_nz_{\rm IR}\gg1$ satisfies the very same equation found in the limit $c_M=0$. This can be readily understood by observing that the dependence on $c_M$ disappears from the equation of motion for $\Sigma=\psi/\chi$, see the first line of~(\ref{eqSigma}), when $x=m_nz\gg c_M$. The massive spectrum is therefore insensitive to the bulk Majorana mass. We numerically verified this statement.

Not much can be inferred analytically regarding the masses and eigenfunctions of states with $m_nz_{\rm IR}\sim1$. Being interested in the lightest eigenstate, which generically belongs to this class, we were therefore forced to solve the eigenvalue problem numerically. We chose $c_D,c_M$, imposed the appropriate boundary conditions on the IR (we will explain shortly what we mean by ``appropriate"), and then numerically evaluated $\chi,\psi$ for a range of $m_n$. The eigenvalues $m_n$ are finally identified by imposing $\psi(x,z_{\rm UV})=\alpha\chi(x,z_{\rm UV})$. Because there is no reason to expect that the physics on the UV scale is chiral (since the bulk itself is not) we will always take $\alpha\neq0$.

We consider a class of IR condition of the form
\ba\label{IRBC}
\psi(x,z_{\rm IR})=\beta\chi(x,z_{\rm IR}).
\ea
It is useful to interpret $\beta$ ($1/\beta$) as a Majorana mass for $\chi$ ($\psi$) localized on the IR brane. A similar interpretation holds for $\alpha$.

Let us first discuss the spectrum corresponding to the limit $\beta\ll1$ (i.e. $\psi(x,z_{\rm IR})=0$). In this case the solution (for a given $m_n$) is completely specified, up to an irrelevant normalization, by setting $\chi(x,z_{\rm IR})$ equal to some arbitrary value, say $\chi(x,z_{\rm IR})=1$. For $c_D>0$ one finds that the function $\psi/\chi$ is essentially flat ($\simeq({c_D-c})/c_M=c_M/(c+c_D)$), up to isolated poles located at the zeros of $\chi$. It follows that the spectrum does not depend on the actual numerical value $\alpha$ defining the UV boundary condition, as long as one stays away from the finely tuned condition $\alpha\approx({c_D-c})/c_M$. This property may be understood by observing that the dual boundary theory associates the bulk fermion to a CFT operator of dimension $d>2$, as argued in Appendix~\ref{app:AdS/CFT}, and that $\alpha$ represents an irrelevant mass deformations.

\begin{figure}%[t] %  figure placement: here, top, bottom, or page
\begin{center}
\includegraphics[width=2.9in]{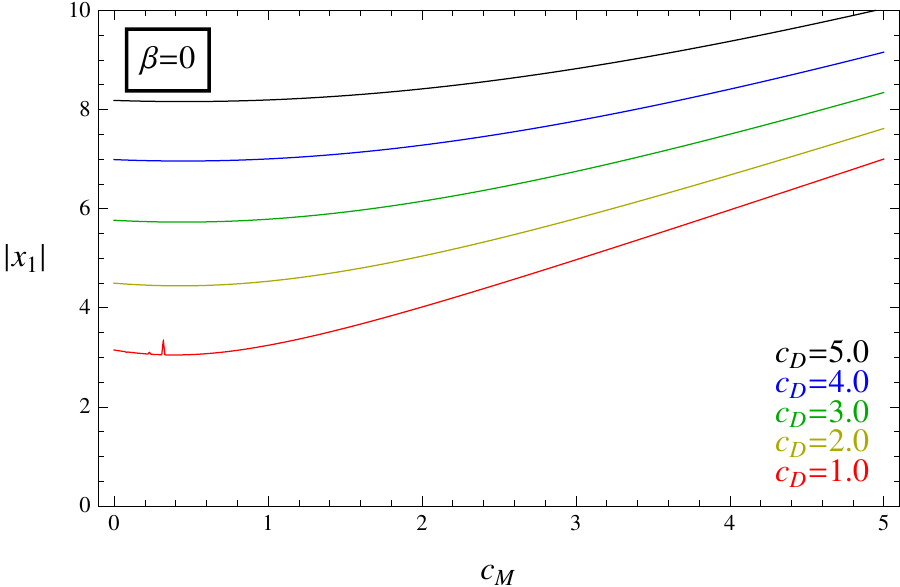}~~~\includegraphics[width=2.9in]{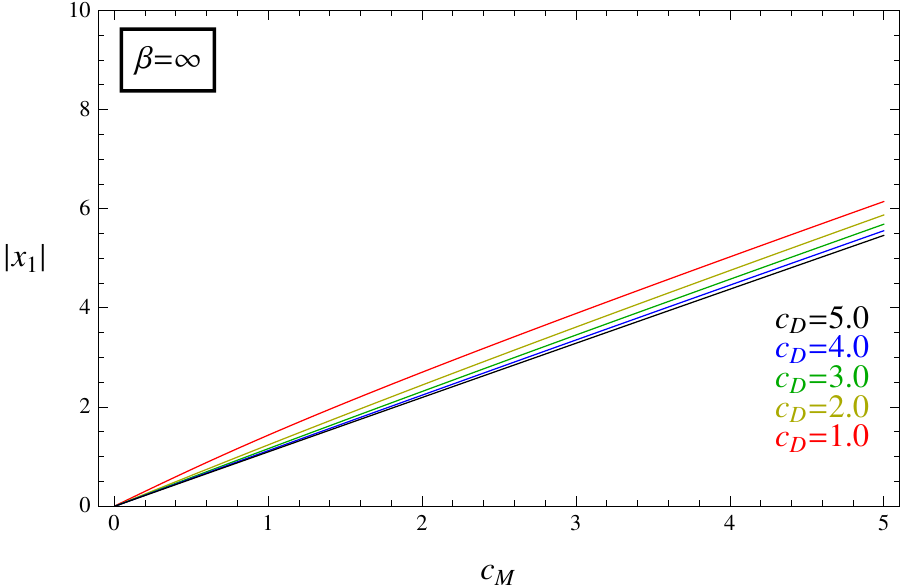}
\\
\includegraphics[width=2.9in]{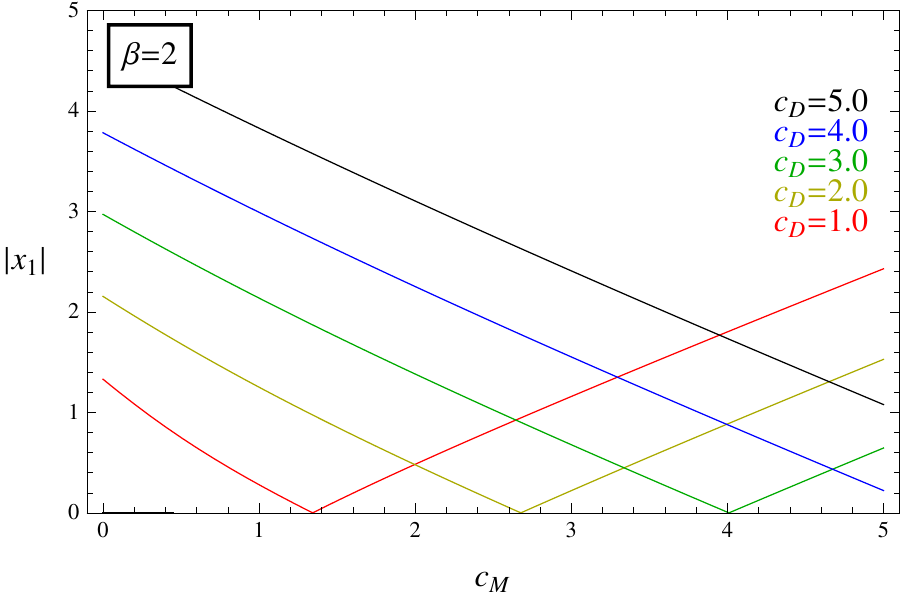}~~~\includegraphics[width=2.9in]{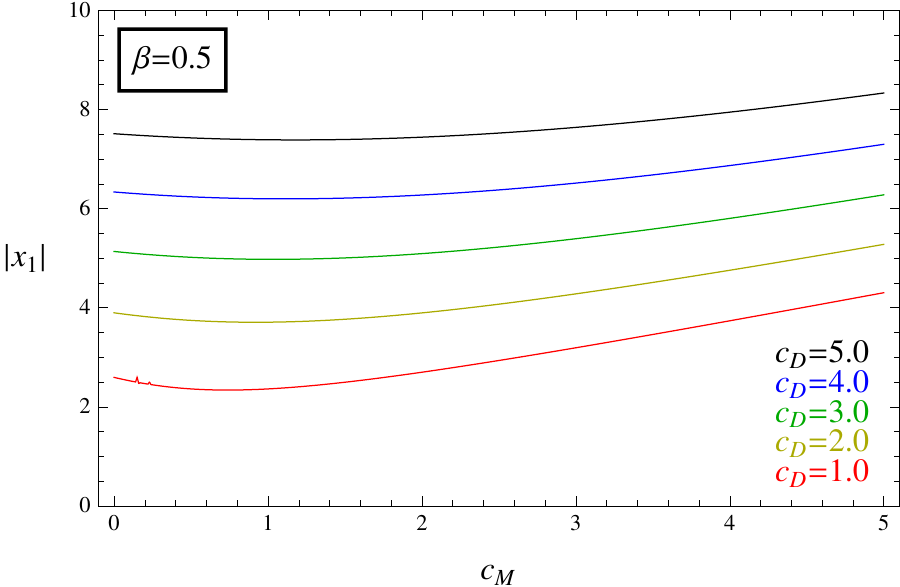}
\caption{\small Mass $|m_1|=|x_1|/z_{\rm IR}$ of the lightest Majorana fermion for various values of $c_{D,M}$. We imposed the boundary conditions $\psi=\beta\chi$ on the IR ($\beta=0,\infty,0.5,2$ in clockwise order starting from top-left), while $\psi=\alpha\chi$ on the UV. For generic $\alpha\neq0$, the UV boundary condition effectively corresponds to $\chi(x,z_{\rm IR})=0$. The DM is parametrically light in the chiral regime $\beta\gg1, c_M\ll1$ (top-right plot), and accidentally light when the parameters get close to the condition~(\ref{light}) (bottom-left plot).
\label{spectrumMajo}}
\end{center}
\end{figure}

In the top-left plot of figure~\ref{spectrumMajo} we show the lightest eigenvalue $|x_1|=|m_1|z_{\rm IR}$ as a function of $c_M$ for some values of $c_D>0$. We took $z_{\rm UV}/z_{\rm IR}=10^{10}$, but the spectrum is basically independent of $z_{\rm UV}$ when $z_{\rm UV}\ll z_{\rm R}$. We varied $c_M$ within $[0,5]$ in linearly spaced steps of $1/500$. Such a fine binning was required since at small $c_{M,D}$ and for such a large hierarchy numerical artifacts tended to appear. We find that $|x_1|$ increases linearly with $c=\sqrt{c_M^2+c_D^2}$, approximately as $|x_1|\sim2.7+c$, so the light fermion mass is always heavier compared to the $c_M=0$ case.

%%%%%%%%%%%%%%%%%%%%%%%%%%%%%%%
%%%%%%%%%%%%%%%%%%%%%%%%%%%%%%

One can repeat a similar exercise for $\chi(x,z_{\rm IR})=0$, or equivalently $\beta\gg1$ (supplemented for example with $\psi(x,z_{\rm IR})=1$). Imposing $\psi(x,z_{\rm UV})=\alpha\chi(x,z_{\rm UV})$, with non-vanishing $\alpha$, also here effectively corresponds to requiring $\chi(x,z_{\rm UV})=0$. Similarly to the previous case, the $|m_n|$s are increasing functions of $c_M$. The crucial difference between the present and the previous IR boundary conditions is that for $\chi(x,z_{\rm IR})=0$ the theory possesses an approximately chiral zero mode localized close to the IR, as it is in the well known $c_M=0$ limit. The behavior of $x_1$ for non-vanishing $c_M$ is shown in the top-right plot of figure~\ref{spectrumMajo}.

It is easy to understand the numerical result. For vanishing Majorana mass $c_M=0$ our boundary conditions admit a zero mode with profile $(\chi_0,\psi_0)=(0,B z^{2+c})$. A small bulk Majorana mass introduces a tiny 4D mass, that we may compute by evaluating the action on $(\chi_0,\psi_0)$. Taking into account the normalization, this procedure gives 
\ba\label{lightcomp}
m_1\approx\frac{\int dz a^5 M_M\psi_0^2}{\int dz a^4\psi_0^2}\approx\frac{2c+1}{2c}\,\frac{c_M}{z_{\rm IR}}.
\ea
We find that this estimate approximates well the exact numerical solution of figure~\ref{spectrumMajo}. In terms of the dual boundary theory, we see that the chiral composite found in the $c_M=0$ limit acquires a Majorana mass of order $c_M$ times the compositeness scale.

It is interesting to discuss in more generality under which conditions we should expect to find a state much lighter than $c_M/z_{\rm IR}$ and/or $c_D/z_{\rm IR}$. For such a state, if it exists, the eigenfunction reduces to the zero mode solution of eq.(\ref{zero}). Imposing our UV boundary condition one finds that this solution survives the IR conditions~(\ref{IRBC}) provided $\beta c_M=c+c_D$, or equivalently
\ba\label{light}
c_M=\frac{2\beta c_D}{\beta^2-1}.
\ea 
For very large $\beta$ we recover the previous result, where the zero lives at low values of $c_M$ (right plot in the top of figure~\ref{spectrumMajo}). For finite $\beta$ we find that a non-negligible portion of parameter space in the neighborhood of the surface~(\ref{light}) has an accidentally light fermion. 

To show this, we plot the mass of the lightest mode for $\beta=2$ in the bottom-left plot of figure~\ref{spectrumMajo}. The zero $x_1=0$ is exactly at the solution of~(\ref{light}). As $\beta$ increases the massless state moves to lower values of $c_M$. On the other hand, for $\beta\to1$ the zero asymptotes to infinity and $|x_1|$ becomes independent of $c_M$. Finally, when $\beta<1$ we approach the scenario $\psi(x,z_{\rm IR})\sim0$ studied at the beginning of this subsection. In this case there is no light state when $c_D>0$ and $c_M>0$, as suggested by~(\ref{light}). This is demonstrated in the bottom-right plot of figure~\ref{spectrumMajo} for $\beta=0.5$. An intuitive interpretation of these results follows by viewing $1/\beta$ as a (IR-localized) Majorana mass for the would-be light chiral mode $\psi_0$.

Finally, we would like to comment on the dual CFT interpretation of a bulk fermion with Majorana mass $M_M\neq0$. The zero mode solution~(\ref{zero}) suggests that the dual picture is the same as in the limit $c_M=0$ (see refs~\cite{Henningson:1998cd}\cite{Mueck:1998iz}\cite{Contino:2004vy}) up to the replacement $c_D\to{\rm sign}(c_D)\sqrt{c_D^2+c_M^2}$. This naive guess turns out to be correct, as shown in Appendix~\ref{app:AdS/CFT}. The result~(\ref{lightcomp}) also shows that $c_M$ provides a measure of the amount of chiral symmetry breaking present in the dual CFT. Consistently, we will see in Appendix~\ref{app:AdS/CFT} that $c_M$ is mapped into a (4D) Majorana mass for both the boundary operator dual to $\Psi$ and its source field.

\section{Relic abundance}
\label{sec:relic}

Unless some additional structure is invoked, $X$ is expected to have unsuppressed couplings to other bulk fields, and hence to be in thermal equilibrium with the thermal bath when the Universe had temperatures $\gg m_X$. In generic 5D scenarios, the density of the neutral fermion $X$ will therefore be set by thermal freeze-out. In this case the relic abundance $\Omega_X$ is essentially determined by the thermally averaged annihilation cross section $\sigma v_{\rm rel}$, which we take to be velocity-independent for later convenience, via the following relation:~\footnote{We assume that no large injection of entropy (from additional particles) occurs at and below freeze-out.}
\ba
\Omega_Xh^2\approx\frac{1.07\times10^{9}}{{\rm GeV}}\frac{x_f}{{g^{1/2}_*}M_{\rm Pl}\,\sigma v_{\rm rel}},
\ea
where $g_*$ is the number of effective degrees of freedom at freeze-out (we take $g_*=107$), $M_{\rm Pl}=1.22\times10^{19}$ GeV, and finally the freeze-out temperature $m_X/x_f$ is defined by:
\ba
x^{1/2}_fe^{x_f}\approx\frac{5}{4\pi^3}\sqrt{\frac{45}{8}}\frac{M_{\rm Pl}m_X\sigma v_{\rm rel}}{g^{1/2}_*}.
\ea
In this section we estimate the main contributions to $\sigma v_{\rm rel}$, and identify the region of the parameter space in which $X$ can account for the totality of the dark matter $\Omega_{\rm DM}h^2=0.12$~\cite{Larson:2010gs}\cite{Ade:2013zuv}. 

We first consider annihilation into final states with other Kaluza-Klein resonances, which are typically accessible for relatively large bulk masses. Annihilation into SM fields is generically inefficient. An exception is the rate $XX\to hh,Z^0Z^0,W^+W^-$, which may account for the observed DM density in models where the couplings between $X$ and the Higgs are unsuppressed. We ignore co-annihilation because its relevance is too model-dependent.

\subsection{Annihilation into heavy resonances}

In any warped model, $X$ interacts with spin-2 gravity excitations, including a potentially light radion $\sigma$. The former interactions have been discussed recently in~\cite{Lee:2013bua}, while the latter in~\cite{Bai:2009ms}. The corresponding annihilation rates are velocity-suppressed and typically smaller than those considered here. 

Other sources of DM annihilation may arise from couplings to new scalars, (spontaneously broken) gauge fields, and fermions. Rather than introducing new fields with the only purpose of getting the right thermal abundance, in this section we will emphasize the importance of the couplings to the Goldberger-Wise (GW) field. 

The GW scalar, here called $\Phi$, is introduced to stabilize the size of the extra dimension~\cite{Goldberger:1999uk}, and is therefore a central ingredient of scenarios with warped extra dimensions. All bulk couplings of $\Phi$ must be small in order to naturally account for the hierarchy $z_{\rm UV}\ll z_{\rm IR}$. However, $\Phi$ has a potential on the IR and UV branes. We therefore assume that the interactions between $X$ and $\Phi$ are suppressed by a small coupling $\epsilon$ in the bulk, but allow $O(1)$ couplings on the IR brane. The rationale behind this expectation is motivated by the dual CFT picture. In the holographic interpretation, $\Phi$ is dual to a nearly marginal deformation of the CFT with scaling dimension $4+\epsilon$, and its bulk couplings of $O(\epsilon)$ control the running of the deformation~\cite{Rattazzi:2000hs}. The CFT eventually flows away from the conformal fixed point, at the IR brane, and all CFT couplings are expected to run fast at that scale. On the gravity side this is precisely the statement that the IR-localized couplings of $\Phi$ are expected to be unsuppressed.

Including all 4D Lorentz bilinears in $\Psi$ on the IR brane, that is $\overline{\Psi}\gamma^5\Psi,\overline{\Psi}\gamma^5\Psi^c,\overline{\Psi}\Psi,\overline{\Psi}\Psi^c$, we will find the couplings
\ba
\frac{y}{2}\overline{X}X\Phi+i\frac{y_5}{2}\overline{X}\gamma^5X\Phi
\ea
between the 4D Majorana field $X$ and the lightest KK mode of the GW scalar, which by an abuse of notation we denote with $\Phi$. It is natural to assume that these couplings are comparable in size to the other couplings of the theory (gauge and Higgs Yukawas), so we take $y,y_5=O(1)$. 

The mass $m_\Phi$ of the lightest KK mode can be estimated without solving Einstein's equations. The reason is that scenarios that address the hierarchy problem have $\epsilon\ll1$, the bulk scalar couplings are small, and the back-reaction of the GW field on the metric is negligible. In the limit $\epsilon\to0$ the heavy KK spectrum is well approximated by the usual equations of motion of a nearly massless 5D scalar. Including an IR-localized mass term $m_{\rm IR}/L$ the eigenvalue problem reads
\ba\label{eps}
m_{\rm IR}J_2(x_n)+x_nJ_1(x_n)=0,
\ea
where $J_\nu(x)$ is the Bessel function of the first kind and order $\nu$. One can explicitly check that, once the coupled gravitational system is solved (see~\cite{Kofman:2004tk}\cite{Elander:2010wd}), the massive KK spectrum is well approximated by the solution of eq.~(\ref{eps}) when $\epsilon\ll1$.

For $|m_{\rm IR}|\gg1$ the KK spectrum is determined by the zeros of $J_2(x_n)$, and $m_\Phi\sim5/z_{\rm IR}$. However, $\Phi$ is usually slightly lighter otherwise. Assuming natural $O(1)$ values for $m_{\rm IR}$ we find that the typical mass of the lightest KK is approximately $m_\Phi\sim4/z_{\rm IR}$. Close to $m_{\rm IR}=-4$ the scalar is anomalously light~\cite{Elander:2010wd}, in some analogy with the accidentally light fermion studied around eq.~(\ref{light}). This mode is UV-localized, and has a wave-function $\sim z^{4+\epsilon}$. Its squared mass may be estimated by evaluating the action on the approximate solution, and reads $(m_\Phi z_{\rm IR})^2\propto4+\epsilon+m_{\rm IR}$. We see that the light state is present only for $m_{\rm IR}>-4$, while for lower IR-localized masses it acquires a tachionic mass and blends with the massive spectrum. 

We now proceed with an estimate of the dominant annihilation modes for our DM candidate $X$.

Annihilation into a pair of $\Phi$, if kinematically open, can proceed via a t-channel exchange of $X$ or via an s-channel pseudo-scalar. In the former case we find 
\ba\label{XXphiphi}
\sigma(X{X}\to \Phi\Phi) v_{\rm rel}=\frac{y^2y_5^2}{8\pi m_X^2}\frac{\left(1-\frac{m_\Phi^2}{m_X^2}\right)^{1/2}}{\left(1-\frac{m_\Phi^2}{2m_X^2}\right)^{2}},
\ea
where we neglected $O(v_{\rm rel}^2)$ terms. The s-channel process involves one power of a scalar trilinear coupling. If this is parametrically smaller than $m_X$, the process will be less efficient. We neglected this contribution for simplicity. The presence of both scalar and pseudo-scalar couplings in~(\ref{XXphiphi}) can be understood by observing that an initial s-wave configuration of identical fermions has $J=0$ and $CP=-1$. For $y_5=0$ annihilation typically proceeds via p-wave (this is the case for final states with a pair of KK gravitons or radions).

When $m_X<m_\Phi$ one might consider a radion and a (axial) $\Phi$ final state:
\ba\label{XXphichi}
\sigma(X{X}\to \Phi\sigma) v_{\rm rel}&=&\frac{y_5^2}{64\pi f^2}\frac{\left(1+\frac{m_\Phi^2-m_\sigma^2}{4m_X^2}\right)^{2}}{\left(1-\frac{m_\Phi^2+m_\sigma^2}{4m_X^2}\right)^{2}}\\\no
&\times&\left(1-\frac{(m_\Phi-m_\sigma)^2}{4m_X^2}\right)^{1/2}\left(1-\frac{(m_\Phi+m_\sigma)^2}{4m_X^2}\right)^{1/2}.
\ea
The pole at $4m_X^2=m_\Phi^2+m_\sigma^2$ is unphysical, but approaches the physical plane when $m_\sigma^2\ll m_\Phi^2$. 

In~(\ref{XXphichi}), the coupling to the radion was simply obtained via the replacement $m_X\to m_X\sigma/f$. In terms of the 5D Planck scale $M$ and curvature $L$, the radion ``decay constant" reads $f=\sqrt{6}(ML)^{3/2}/z_{\rm IR}$ (see for example~\cite{Csaki:2000zn} and~\cite{Rattazzi:2000hs}). If (\ref{XXphichi}) is not accessible either, $XX\to2\sigma$ might be open. This possibility will be considered below.

We next consider annihilation into exotic fermions $Q$. For example, the $Q$s could be identified with the first KK mode of the bulk SM fermions, which will eventually decay into SM fermions and a Higgs or a gauge bosons. The processes $X{X}\to\overline{Q}Q$ can be induced by the exchange of some bulk scalar coupled to $\overline{\Psi}\Psi^c,\overline{\Psi}\Psi$ and $\overline{Q}Q$. The most minimal scenario makes again use of the GW scalar, that is assumed to have a (IR-localized) coupling to $Q$ as well. Writing this latter in 4D notation as $y_Q\overline{Q}Q\Phi$ we find
\ba\label{XXQQbar}
\sigma(X{X}\to\overline{Q}Q) v_{\rm rel}=N_Q\frac{y_Q^2y_5^2}{2\pi}\frac{m_X^2}{\left|4m_X^2-m_\Phi^2+i\Gamma_\Phi m_\Phi\right|^2}\left(1-\frac{m_Q^2}{m_X^2}\right)^{3/2},
\ea
where $N_Q$ is the number of kinematically accessible $Q$s. We ignore the scalar-scalar coupling $\overline{X}X\overline{Q}Q$ because it leads to a velocity-suppressed rate.

For DM masses below $m_Q$ the process $XX\to\overline{Q}q_{L,R}$, also mediated by and s-channel $\Phi$, is active (the same annihilation mode, though mediated by a higher dimensional operator, was also discussed in~\cite{Ponton:2008zv} for scalar DM). The rate for the production of a single chirality of $q$ is 
\ba\label{XXQt}
\sigma(X{X}\to\overline{Q}q_{L,R}) v_{\rm rel}&=&\epsilon_q^2N_q\frac{y_Q^2y_5^2}{4\pi}\frac{m_X^2}{\left|4m_X^2-m_\Phi^2+i\Gamma_\Phi m_\Phi\right|^2}\left(1-\frac{m^2_Q}{4m_X^2}\right)^2,
\ea
where we safely neglected terms of $O(m_q^2/m_Q^2)$. Here $N_q$ is the number of allowed fermionic pairs and $\epsilon_q$ measures the overlap of the profile of the SM fermion on the IR, which is very sensitive to its localization along the extra dimension. As a realistic estimate one could identify $q$ with the top quark, in which case $N_q=3$ and $\epsilon_q^2\gtrsim1/4\pi$.

\begin{figure}%[t] %  figure placement: here, top, bottom, or page
\begin{center}
\includegraphics[width=2.7in]{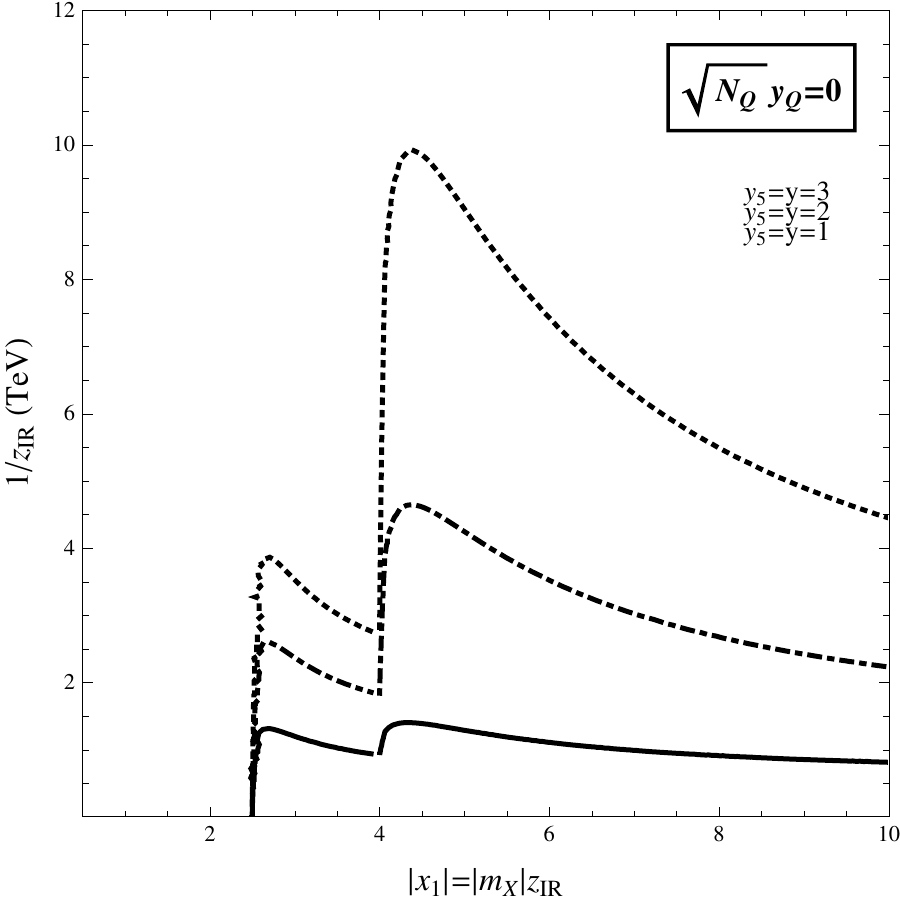}~~~~~~~~~~~~\includegraphics[width=2.7in]{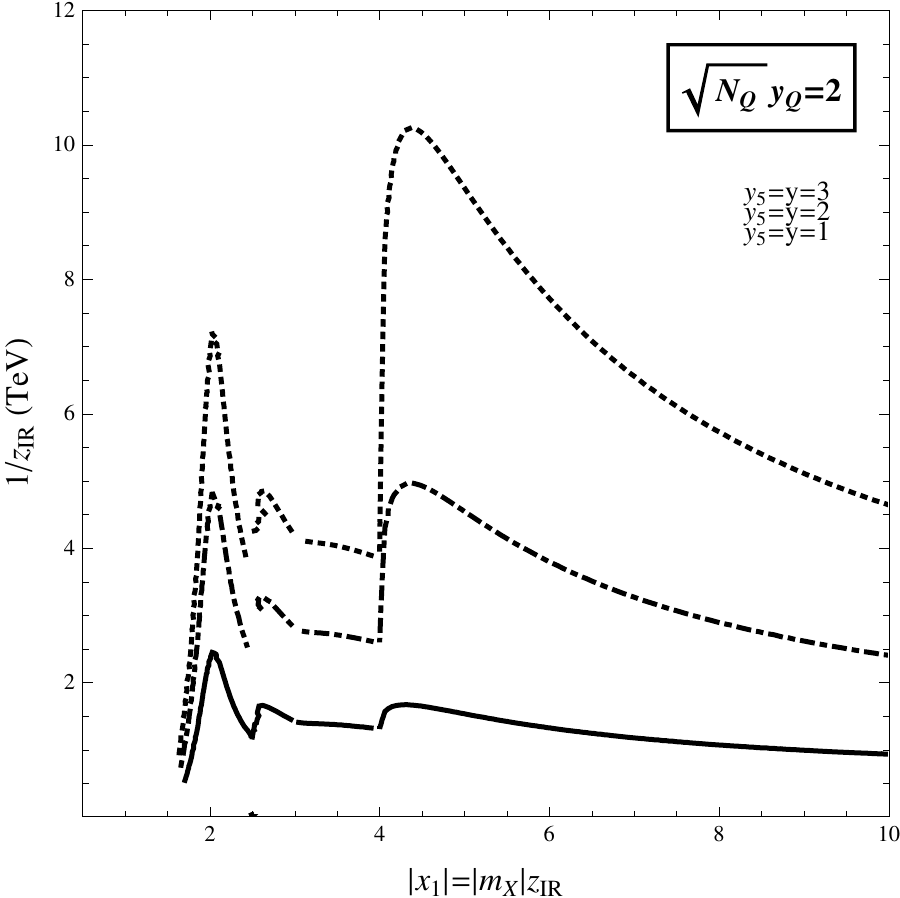}
\caption{\small Parameter space $(|x_1|, z_{\rm IR})$ satisfying the relic density constraint $\Omega_Xh^2=0.12$~\cite{Larson:2010gs}\cite{Ade:2013zuv} for different values of the couplings in eqs.~(\ref{XXphiphi}),~(\ref{XXphichi}),~(\ref{XXQQbar}),~(\ref{XXQt}). The plot on the left has $y_Q=0$ whereas on the right $\sqrt{N_Q}y_Q=2$ and $\epsilon_q^2N_qy_Q^2=0.3N_Qy_Q^2$. The three lines correspond, from bottom up, to $y,y_5=1$ (solid), $=2$ (dot-dashed), $=3$ (dotted). See text for more details. 
\label{annM}}
\end{center}
\end{figure}

In figure~\ref{annM} we present the constraint $\Omega_Xh^2=0.12$ in the plane $(|x_1|=m_Xz_{\rm IR},z_{\rm IR})$ for $y=y_5=1,2,3$ (solid lines from bottom up). In the left plot only the rates in (\ref{XXphiphi}), (\ref{XXphichi}) contribute, while on the right eqs.~(\ref{XXQQbar}) and~(\ref{XXQt}) were also turned on with $\sqrt{N_Q}y_Q=2$ and $\epsilon_q^2N_q=0.3 N_Q$. We assumed the lightest KK of the GW field has a mass $m_\Phi=4/z_{\rm IR}$, which is approximately what one finds in models with an IR-localized mass of order unity. We also used $\Gamma_\Phi/m_\Phi=y_Q^2N_Q/8\pi$, and further took $m_Q=3/z_{\rm IR}$ as a conservative estimate for the mass of the heavy partners of the light SM quarks. Finally, the radion mass is taken to be $m_\sigma=1/z_{\rm IR}$ and $f=\sqrt{6}/z_{\rm IR}$.

%%%%%%%%%%%%%%%%%%%%%%%%%%%%%%%
%%%%%%%%%%%%%%%%%%%%%%%%%%%%%%

For $m_Q<m_X<m_\Phi$ the major contributions are from $XX\to\Phi\sigma$ and $XX\to QQ$, and which one dominates depends on whether $2N_Qy_Q^2f^2/m_Q^2$ is bigger or smaller than one. With our choice of parameters~(\ref{XXQQbar}) gives a non-negligible contribution (compare the two plots in this mass range). When the DM is above $m_\Phi$, the very efficient channel $XX\to\Phi\Phi$ opens, and the relic constraint simply reads $m_X\sim yy_5\times4.5$ TeV. Increasing further the DM mass, while keeping the couplings fixed, eq.~(\ref{XXphichi}) will eventually win, and the constraint $\Omega_Xh^2=0.12$ becomes $f\sim y_5\times 1.7$ TeV.

Realistic warped dimensional models must have $2.4/z_{\rm IR}\gtrsim4-5$ TeV in order to be compatible with the electroweak data. It is interesting to observe that DM also favors $1/z_{\rm IR}$ above the TeV, provided $m_X\gtrsim O(1)/z_{\rm IR}$. Our analysis in section~\ref{sec:spectrum} shows that ``neutralino" masses of this order are generic away from the chiral limit $c_M\ll1$, see figure~\ref{spectrumMajo}.

We emphasized the central role of the Goldberger-Wise field $\Phi$ and its (IR-localized) couplings to DM {\emph{bilinears}}, which are responsible for triggering the important modes $XX\to\Phi\Phi,\Phi\sigma$ and are generic in this framework. The existence of a neutral scalar $\Phi$ is a welcome feature of the Randall-Sundrum scenario. In theories without neutral spin-0 multiplets the neutralino couples {\emph{linearly}} to the SM via exotic states. In those cases annihilation proceeds via the virtual exchange of the new states which, as opposed to $\Phi$, carry the accidental ${\mathbb{Z}}_2$ parity and must be heavier by construction.

\subsection{Annihilation into SM fields (and the radion)}
\label{sec:lightDM}

The DM can be parametrically lighter than $O(1)/z_{\rm IR}$ in the nearly chiral limit $c_M\ll c_D$, and may also be {\emph{accidentally}} light when approaching the condition~(\ref{light}) (see the bottom-left plot in figure~\ref{spectrumMajo}). In those cases annihilation into a radion pair as well as SM particles, if allowed, will control the relic abundance. 

The former gives~\footnote{Our result is a factor $4/3$ larger than that of~\cite{Bai:2009ms}.}
\ba\label{XXsigmasigma}
\sigma(X{X}\to \sigma\sigma) v_{\rm rel}=\frac{m_X^2}{32\pi f^4}v_{\rm rel}^2\frac{\left(1-\frac{m_\sigma^2}{m_X^2}\right)^{1/2}}{\left(1-\frac{m_\sigma^2}{2m_X^2}\right)^{2}}.
\ea
In the optimal limit $m_\sigma\ll m_X$, taking $m_X=|x_1|/z_{\rm IR}$ and conservatively assuming $f=\sqrt{6}/z_{\rm IR}$, we find that this cross section is of the right oder of magnitude at freeze-out provided $|x_1|z_{\rm IR}=O(1/100$ GeV$)$. In a realistic theory with $2.4/z_{\rm IR}>4$ TeV the rate is irrelevant unless $|x_1|=O(10)$, so large that the annihilation modes of the previous section are open. We conclude that the mode $XX\to\sigma\sigma$ is not important.

Final states with SM fermions described by dimension-6 operators $\epsilon^2_q\overline{X}\gamma^\mu\gamma^5 X\overline{q}\gamma^\mu q/2f^2$ lead to $\sigma(X{X}\to q\overline{q}) v_{\rm rel}=\epsilon_q^4{N_cm_X^2}v_{\rm rel}^2/{6\pi f^4}$, where we neglected the SM fermion mass. Even for a maximally composite SM fermion, $\epsilon_q\sim1$, this is negligible at freeze-out unless $|x_1|\gtrsim3$, in which case the s-wave processes discussed in the previous subsection dominate.

Next we consider annihilation into the SM Higgs. The operator $\overline{X}\gamma^\mu\gamma^5 X iH^\dagger\overleftrightarrow{D_\mu}H$ mediates a p-wave process, and arguments similar to $XX\to\sigma\sigma,q\overline{q}$ apply. (Furthermore, this rate may be suppressed by invoking a custodial $SU(2)_c$ and appropriate charges for $X$). 

In models with light $X$, and no new exotic light particles, the main contribution to DM annihilation at freeze-out will come from
\ba\label{XH}
\frac{y_H^2}{2\Lambda}\overline{X}XH^\dagger H+\frac{y_{H5}^2}{2\Lambda}i\overline{X}\gamma^5XH^\dagger H.
\ea
This might arise from bulk or IR-localized Yukawa couplings of $\Psi$ and $\Psi^c$ to some heavy fermion doublet of mass $\Lambda$ and the Higgs, or the exchange of some scalar coupled to $H^\dagger H$ and $\Psi$ bilinears. The operators~(\ref{XH}) mediate annihilation into Higgses, gauge bosons, and SM fermions. We neglect final states with SM fermions because suppressed by $m_t^2/m_X^2$ compared to the others. The rate for $XX\to hh$ reads:
\ba
\sigma(X{X}\to hh) v_{\rm rel}&=&\frac{1}{\Lambda^2}\left[y_{H5}^4+y_{H}^4\left(1-\frac{4m_X^2}{s}\right)\right]\\\no
&\times&\frac{s}{128\pi m_X^2}\left(1-\frac{4m_h^2}{s}\right)^{1/2},
\ea
and similarly for $XX\to VV=W^+W^-,Z^0Z^0$:
\ba
\sigma(X{X}\to VV) v_{\rm rel}&=&\frac{1}{\Lambda^2}\left[y_{H5}^4+y_{H}^4\left(1-\frac{4m_X^2}{s}\right)\right]\\\no
&\times&\frac{c_Vs}{128\pi m_X^2}\left(\frac{s}{s-m_h^2}\right)^2\left(1+\frac{8m_V^2}{s}\right)\left(1-\frac{4m_V^2}{s}\right)^{1/2},
\ea
with $c_V=1(2)$ for $V=Z^0$ ($W^\pm$). The rate $\propto y_{H5}^4$ proceeds via an s-wave, and provides the dominant contribution. At center of mass energies above $m^2_h$, the total cross section quickly asymptotes to the value $\sigma(XX\to H^\dagger H)v_{\rm rel}=[y_{H5}^4+(y_{H}^4+y_{H5}^4)v_{\rm rel}^2/4]/8\pi\Lambda^2$ predicted by the equivalence theorem. The observed DM density is then obtained either for $y_{H5}^2\sim\Lambda/4.6$ TeV, or $y_{H}^2\sim\Lambda/(700-800$ GeV$)$ when $y_{H5}$ is small. In this latter case a largish $y_{H}$ is needed in realistic scenarios with $\Lambda\gtrsim$ few TeV, and the correction to the DM mass induced by the operators~(\ref{XH}) may be sizable.

If $H,X$ are generic resonances, naive dimensional analysis suggests that $y_{H,H5}$ are of the order of a typical coupling between KK states. The coefficients of (\ref{XH}) are instead suppressed by a chiral factor $m_X/\Lambda\ll1$ when the DM is parametrically lighter than the typical KK scale $\Lambda$. An additional suppression of order ${\delta m_h^2}/{\Lambda^2}$ is present in models where the Higgs is a pseudo-Nambu Goldstone mode, with $\delta m_h^2\ll\Lambda^2$ the size of typical corrections of the Higgs mass. In all these latter scenarios, obtaining the right relic abundance is challenging for light $X$.

\section{Experimental signatures}
\label{sec:signatures}

DM couplings to the SM can be constrained by collider experiments. Currently, searches of missing energy signatures performed by the CMS and ATLAS collaborations place bounds of the order of a few hundred GeV on the new physics scale (see for instance~\cite{Chatrchyan:2012me} and~\cite{Aad:2013oja}). These bounds quickly drop as the DM mass gets in the TeV range, and are very sensitive to the quantum numbers and masses of the particles mediating the DM-SM interactions. 

A more efficient and model-independent probe of the present framework is provided by direct (and possibly indirect) DM searches.

\subsection{Direct searches}

The direct detection signatures of $X$ are similar to those of a supersymmetric heavy neutralino. In this subsection we outline the main constraints, while a more detailed analysis will be presented in a forthcoming paper.

Among the most relevant couplings are those given in eq.~(\ref{XH}) (${\cal O}_{Xy}=\overline{X}Xy\overline{q}Hq/\Lambda f^2$ leads to comparable effects). We can neglect the CP-odd operator because its contribution to the elastic scattering rate is suppressed by the small DM velocity in the halo, $v_{\rm DM}\sim10^{-3}$, so we are left with
\ba
{\cal O}_{XH}=\frac{y_H^2}{2\Lambda}\overline{X}XH^\dagger H. 
\ea
Once the Higgs is integrated out, ${\cal O}_{XH}$ generates a 4-fermion interaction for quarks and DM. This is matched onto a coupling to the nucleon $N$ via standard methods, $\langle N|\sum_{q}m_q\overline{q}q|N\rangle\to f_N\overline{N}N$, with $q$ running over all 6 SM flavors and $|N\rangle$ denoting the nucleon one-particle state. As a representative value for $f_N$ we make an average of the estimates quoted in refs.~\cite{matrix}. This turns out to be $f_n,f_p\approx380$ MeV for the neutron and proton. It is now a trivial exercise to derive the differential rate for scattering of DM and target nuclei of mass $m_T$ (assumed to be composed of a single species):
\ba
\frac{d\sigma_T}{dE_R}=\frac{\mu_T^2}{\pi m_h^4}\left(\frac{y_H^2}{\Lambda}\right)^2\left[Zf_p+(A-Z)f_n\right]^2\frac{|F(E_R)|^2}{E_R^{\rm max}}.
\ea
Here $\mu_T=m_Xm_T/(m_X+m_T)$ is the reduced mass of the DM-target system, $E^{\rm max}_R$ is the maximum target recoil energy, and $F(E_R)$ a nuclear form factor. To derive our bounds we use the LUX~\cite{Akerib:2013tjd} and 2012 XENON100 data~\cite{Aprile:2012nq}, which provide the most stringent constraints for heavy DM (for comparison, bounds from the CDMS collaboration can be found here~\cite{Ahmed:2009zw}). For $m_X>m_T$ the total event rate depends on the DM mass mainly via the local DM density, which we take to be $0.3$ GeV cm$^{-3}/m_X$. In this mass range we can therefore write the $90\%$ CL bound simply as
\ba\label{XH'}
\Lambda\gtrsim(1.9-2.6)~{\rm TeV}\times y_{H}^2\sqrt{\frac{\rm TeV}{m_X}}.
\ea 
This roughly corresponds to a bound on the DM nucleon cross section of $\frac{\mu_N^2}{\pi m_h^4}\left(\frac{y_H^2f_N}{\Lambda}\right)^2\lesssim(1-2)\times10^{-44}$ cm$^2\times m_X/$TeV.

As discussed in the previous section, in theories with light DM ${\cal O}_{XH}$ might actually dominate at freeze-out. Assuming $y_{H5}$ is negligible, this assumption basically fixes the coefficient of ${\cal O}_{XH}$. The direct detection bound then translates into a strong lower bound on the DM mass, $m_X\gtrsim6-14$ TeV. On the other hand, if we assume $y_H\sim y_{H5}$, then lower values of the couplings are sufficient to saturate the DM density. The bound is now rather mild, $m_X\gtrsim200-300$ GeV, and a much larger portion of parameter space can be compatible with data. 

XENON1T is expected to improve upon XENON100 by a factor $\sim100$ in sensitivity~\cite{XENON1T}, and will hence push the bounds on $m_X$ up by approximately two orders of magnitude. Imposing the consistency condition $m_X<\Lambda$, we see that most of the parameter space compatible with the perturbative bound $y_{H,H5}^2\lesssim4\pi$ will soon be explored.

In models where the Higgs is a pseudo-Nambu Goldstone boson ${\cal O}_{XH}$ is suppressed, and other interactions may be more relevant. An important operator contributing to spin-independent direct detection signatures is:
\ba
{\cal O}_{XG}\equiv\frac{g_s^2}{2\Lambda^3}\overline{X}XG_{\mu\nu}G^{\mu\nu}.
\ea
To estimate the effect of ${\cal O}_{XG}$ we can match it onto the nucleon matrix elements according to the anomaly relation~\cite{Shifman:1978zn} $\frac{\beta_s}{2g_s}G_{\mu\nu}G^{\mu\nu}\to m_N f_{TG}^{(N)}\overline{N}N$. For concreteness we identify the QCD beta function $\beta_s$ with the 1-loop result $\beta_s=9g_s^3/16\pi^2$. The differential rate for DM scattering off target nuclei is
\ba
\frac{d\sigma_T}{dE_R}=\frac{\mu_{T}^2}{\pi}\left(f_{TG}^{(N)}\frac{32\pi^2m_NA}{9\Lambda^3}\right)^2\frac{|F(E_R)|^2}{E_R^{\rm max}},
\ea
with $|f_{TG}^{(N)}|\sim1$. An analysis analogous to that leading to~(\ref{XH}) here gives the $90\%$ CL bound
\ba\label{XG}
\Lambda\gtrsim1.5~{\rm TeV}\times|f_{TG}^{(n)}|^{1/3}\left(\frac{\rm TeV}{m_X}\right)^{1/6}. 
\ea
Note that ${\cal O}_{XG}$ is dimension-7, so the dependence of the bound on the mass is much weaker than for ${\cal O}_{XH}$.

There are other SM operators that a DM pair can couple to. However, these lead to velocity-suppressed or spin-dependent rates, and often their relevance is more model-dependent. In all these cases the experimental bounds are weaker than those discussed for ${\cal O}_{XG}$ and ${\cal O}_{XH,Xy}$. Moreover, a light radion can enhance the coefficients of ${\cal O}_{XH,Xy,XG}$ by a factor $\sim(\Lambda/m_\sigma)^2$, and so the direct detection bounds discussed above are expected to be stronger in models with $m_\sigma\ll\Lambda$.

\subsection{Indirect searches}

DM annihilation at our times generically occurs at a rate comparable to the freeze-out value $\sim2\times10^{-26}$ cm$^3$/s. The observed fluxes of anti-matter and $\gamma$-rays are currently not constraining as soon as $m_X$ is above a few hundred GeV. For TeV candidates, DM decay is potentially more relevant.

While $U(1)_B\times U(1)_L$ must be conserved in the bulk (and on the IR brane), there is no reason to expect that physics at the UV cutoff respects it. Typically, these new effects will not cause any harm to the proton lifetime and neutrino masses as long as $1/z_{\rm UV}\gtrsim10^{16}$ GeV. However, even such a large UV cutoff might significantly alter the DM phenomenology. In particular, DM decay may occur at too high rates.

The dominant UV-localized couplings for Majorana DM are expected to be $\overline{X}\ell H$ in minimal scenarios ($\ell$ is the SM lepton doublet) and $\overline{X}N$ in models with a right-handed neutrino $N$. Both interactions will trigger DM decay, and requiring that the DM is cosmologically stable implies a very strong bound on the couplings of these operators. The resulting constraint is so strong that $X$ cannot appreciably affect neutrino masses.~\footnote{$X$ cannot be thought of as a right-handed neutrino because it does not have lepton charge.} For example, denoting by $y_\ell$ the coupling of $\overline{X}\ell H$, and assuming $m_X\gg m_\nu$, we have $\delta m_\nu\sim y_\ell^2 v^2/m_X$ and a decay rate for $X$ of order $\Gamma(X\to H\ell)\sim y_\ell^2m_X/4\pi\sim \delta m_\nu m_X^2/4\pi v^2$. A DM lifetime greater than the age of the Universe then translates into a negligible correction to the neutrino mass unless $m_X$ is much smaller than the neutrinos themselves.

In our model $y_\ell$ is generated by the overlap of the DM profile and $H\ell$, and can be naturally small because $X$ and $H$ are localized towards the IR brane while $\ell$ presumably lives closer to the UV brane. In 4D language we expect $y_\ell\sim(z_{\rm UV}/z_{\rm IR})^{d+d_H-5/2}$, with $d$ ($d_H$) the scaling dimension of the CFT operator dual to $X$ (the Higgs $H$). Current bounds from searches of anti-matter and gamma ray fluxes approximately read $1/\Gamma(X\to H\ell)\gtrsim10^{27}$ for $X$ above the TeV (see for instance~\cite{Ibarra:2013cra} for a recent summary). With a UV cutoff close to the Planck scale, and recalling that a solution of the hierarchy problem demands $d_H\gtrsim2$, we find that this translates into $d\gtrsim2$. As shown in Appendix~\ref{app:AdS/CFT} this condition can easily be satisfied with natural bulk masses $M_{D,M}L=O(1)$.

The phenomenology of scenarios with light right-handed neutrinos is richer, but these conclusions are qualitatively unchanged.

\section{Conclusions}
\label{sec:conclusions}

DM is a spin-off of generic models that (approximately) respect the SM baryon and lepton symmetries. Warped extra dimensions is no exception. Realistic Randall-Sundrum scenarios typically have exotic particles satisfying either condition~(\ref{stability1}) or (\ref{stability2}): if these particles are also color and electric neutral, they are automatically DM candidates. 

In this paper we presented a model-independent analysis of warped scenarios with a neutral fermion candidate $X$. We found that this particle can naturally constitute the totality of DM; no additional structure is needed beyond $X$ itself and all couplings compatible with the symmetries. 

The bulk of the parameter space predicts a DM mass of order a few times the inverse of the compactification scale, $1/z_{\rm IR}$. For such values, annihilation into other KK modes (such as GW fields $\Phi$, radions $\sigma$, partners of the SM fermions $Q$, etc.) is possible, and the observed relic density is obtained for $O(1)$ couplings and $1/z_{\rm IR}\gtrsim$ TeV.

DM could be lighter than $O(1)/z_{\rm IR}$. In this case a consistent picture for DM may be obtained if $X$ has unsuppressed couplings with the SM Higgs boson, which can occur in models where the Higgs boson is accidentally light. Direct DM searches provide stringent bounds on this scenario, and will soon be able to probe a significant portion of its parameter space.

\acknowledgments

We thank Kaustubh Agashe for discussions and Marco Nardecchia for comments. We acknowledge the Kavli Institute for Theoretical Physics, where this project was initiated, and the Galileo Galilei Institute, where part of this work was performed. This work was supported in part by the NSF Grant No. PHY-0968854, No. PHY-0910467, by the Maryland Center for Fundamental Physics, and by the National Science Foundation under Grant No. PHY11-25915.

\appendix

\section{Neutral fermions in a Warped 5D geometry}
\label{app:spectrum}

Consider the metric
\ba
ds^2=G_{MN}dx^Mdx^N=a^2(z)(dx^\mu dx_\mu-dz^2),
\ea
with $M=0,1,2,3,5$ and $x^M=(x^\mu,z)$. The most general, manifestly hermitian 5D action for a neutral fermion $\Psi$ quadratic in the field is
\ba\label{action}
{\cal S}&=&\int d^5x\sqrt{-G}\left[\frac{i}{2}\bar\Psi e^M_a\Gamma^aD_M\Psi-\frac{i}{2}(D_M\Psi)^\dagger\gamma^0e^M_a\gamma^a\Psi\right.\\\no
&-&\left.M_D\overline{\Psi}\Psi-\frac{M_M}{2}\overline{\Psi}\Psi^c-\frac{M^*_M}{2}\overline{\Psi^c}\Psi\right].
\ea
Note that the phase of $M_M$ can be removed by an appropriate re-definition of $\Psi$ (the results of this section are however general). Here $a=0,1,2,3,5$ are indices in the tangent space, $\Gamma^\mu=\gamma^\mu$ and $\Gamma^5=-i\gamma^5$ are defined in Weyl's basis, the funfbein is $e^M_a=\delta^M_a/a$, and finally
\ba
D_\mu=\partial_\mu-\frac{i}{2}\frac{\partial_za}{a}\gamma_\mu\gamma^5~~~~~~~~~~~~D_5=\partial_z.
\ea
We also defined
\ba
\Psi^c\equiv C_5\Psi^*,
\ea
where the 5D charge conjugation matrix $C_5$ satisfies $C_5\Gamma^aC_5^{-1}=+(\Gamma^a)^*$ (in 5D, the alternative possibility $C'_5\Gamma^a{C'_5}^{-1}=-(\Gamma^a)^*$ is not compatible with $\Pi_a\Gamma^a=-1$). This relation ensures that $\Psi^c,\Phi$ transform the same way under the 5D Lorentz group.

We employ the convention $C_5=\Gamma^5\Gamma^2$. Note that $C_5C_5^*=-1$ implies that the condition $\Psi=\Psi^c$ has only the trivial solution (see also the explicit expression in~(\ref{expl})).

Varying the action we obtain
\ba
\delta{\cal S}&=&\int d^5x\left\{\left[a^4\delta\overline{\Psi} {\cal D}\Psi+a^4({\cal D}\Psi)^\dagger\gamma^0\delta\Psi\right]+\frac{1}{2}\left[a^4\overline{\Psi}\gamma^5\delta\Psi-a^4\delta\overline{\Psi}\gamma^5\Psi\right]'\right\},
\ea
where
\ba\label{EOM}
{\cal D}\Psi=i\gamma^\mu\partial_\mu\Psi+\gamma^5\Psi'+2\frac{\partial_za}{a}\gamma^5\Psi-aM_D\Psi-aM_M\Psi^c.
\ea
The equations of motion simply read ${\cal D}\Psi=0$.

In terms of the left-handed fermions $\chi,\psi$ introduced in eq.~(\ref{expl}) the equations of motion can be written as
\ba\label{5DEOM}
\Delta\left( \begin{array}{c}  
\chi\\
\psi
\end{array}\right)
\equiv\left( \begin{array}{c}  
+\psi'+2\frac{\partial_z a}{a}\psi-aM_D\phi-aM_M^*\chi\\
-\chi'-2\frac{\partial_z a}{a}\chi-aM_D\chi+aM_M\psi
\end{array}\right)
=-i\sigma^\mu\epsilon\partial_\mu
\left( \begin{array}{c}  
\chi\\
\psi
\end{array}\right)^*.
\ea
The boundary conditions are $\frac{a^4}{2}(\psi^t\epsilon\delta\chi-\delta\psi^t\epsilon\chi+{\rm hc})=0$, up to possible additional boundary terms. As usual, the system~(\ref{5DEOM}) may be turned into two decoupled second order differential equations for $\chi,\psi$. These are the same as found for $c_M=0$ except for corrections of order $c_M/(x\pm c_M)$ and $c_M/c_D$. Unfortunately there is no analytic solution when $c_M,c_D\neq0$.

It is straightforward to prove that the linear map $\phi\to(\Delta\phi)^*$ is hermitian in the space of 2-component vectors $\phi$ with scalar product $\langle\phi_1|\phi_2\rangle\equiv\frac{1}{2}\int dz\;a^4\,\phi_1^\dagger\phi_2+{\rm hc}$ and boundary conditions $\frac{a^4}{2}\phi_1^\dagger i\tau^2\phi^*_2+{\rm hc}=0$. The set 
\ba
\phi_n(z)=
\left( \begin{array}{c}  \hat\chi_n(z) \\
\hat\psi_n(z)
\end{array}\right)
\ea
of eigenvectors $\phi_n$ satisfying $\Delta\phi_n=m_n\phi^*_n$, with $m_n$ real, is therefore complete and orthogonal, and can be used to perform a standard Kaluza-Klein reduction. Working with normalized eigenfunctions, $\langle\phi_n|\phi_m\rangle=\delta_{mn}$, we have 
\ba
\left( \begin{array}{c}  \chi(x,z) \\
\psi(x,z)
\end{array}\right)=\sum_n\phi_n(z)N_n(x),
\ea
The 5D equations of motion reduce to $-i\sigma^\mu\partial_\mu(\epsilon N_n^*)=m_nN_n$ in terms of the left-handed fields $N(x)$. Consistently, the action~(\ref{action}) becomes
\ba
{\cal S}=\sum_n\int d^4\left[iN_n^\dagger\bar{\sigma}^\mu\partial_\mu N_n-\frac{m_n}{2}\left(N_n^t\epsilon N_n+{\rm hc}\right)\right].
\ea
The spectrum is discussed in section~\ref{sec:spectrum}. Throughout the paper we use the 4-component notation $X^t=(N_1,\epsilon N_1^*)$, such that $X$ is a 4-component self-conjugate field (in 4D this means $X=C_4X^*$, where $C_4\equiv-i\gamma^2$ satisfies $C_4\gamma^\mu C_4^{-1}=-(\gamma^\mu)^*$ and $C_4C_4^*=1$).

A convenient choice of field basis is one with real $M_M$. With this choice the eigenfunctions $\phi_n$ can be taken to be real, too. This is the basis chosen in our numerical analysis of section~\ref{sec:spectrum}, and also adopted in Appendix~\ref{app:AdS/CFT}.

\section{Holography with $M_{D,M}\neq0$}
\label{app:AdS/CFT}

According to the AdS/CFT correspondence, $\Psi$ is dual to an operator ${\cal O}$ of the boundary CFT. We will see that $M_M$ enters not only in the anomalous dimension of ${\cal O}$, but also as an explicit breaking of the chiral symmetry of the boundary theory.

In order to proceed with the correspondence we identify $\chi(x,z_{\rm UV})\equiv J(x)$ as the source field. This just amounts to a redefinition $N(p)\to J(p)/\chi_p(z_{\rm UV})$ in our KK decomposition in Fourier space
\ba
\chi(p,z)=\frac{\chi_p(z)}{\chi_p(z_{\rm UV})}J(p)~~~~~~~~~~\psi(p,z)=\frac{\psi_p(z)}{\chi_p(z_{\rm UV})}J(p).
\ea
The UV boundary conditions are therefore $\delta J=0$, while $\psi$ is in general free to vary. These conditions are automatically satisfied if we replace the action in~(\ref{action}) with~\cite{Contino:2004vy}
\ba
{\cal S}\to{\cal S}+{\cal S}_{\rm UV},~~~~~~~~~{\cal S}_{\rm UV}=\frac{1}{2}\int d^4 x\sqrt{-g_{\rm ind}}~\overline{\Psi}\Psi.
\ea
where $\sqrt{-g_{\rm ind}}=a^4(z_{\rm UV})$. The BCs on the interior of $AdS$ are irrelevant for the present discussion, since we will be interested in momentum scales of order $pz_{\rm IR}\gg1$. Evaluating the total action on the solutions of the equations of motion we get
\ba
{\cal S}^{\rm on-shell}={\cal S}_{\rm UV}^{\rm on-shell}=a^4(z_{\rm UV})\int\frac{d^4p}{(2\pi)^4}J^\dagger(p)\frac{\bar\sigma^\mu p_\mu}{|p|}J(p)\Sigma(pz_{\rm UV}),
\ea
with 
\ba
\Sigma(pz)\equiv\frac{\psi_p(z)}{\chi_p(z)}.
\ea
The 2-point function of the dual operator ${\cal O}$ is given by $\langle{\cal O}\overline{\cal O}\rangle=ia^4(z_{\rm UV})\Sigma(pz_{\rm UV})\bar\sigma^\mu p_\mu/|p|$. We expect that $\Sigma$ can be written as the sum of local and non-local terms $\Sigma=\Sigma_{\rm poly}+\delta\Sigma$, with $\Sigma_{\rm poly}=\sum_na_nx^n$ a polynomial in $x$. The leading non-local part in the limit $x\to0$ (note that $x=pz$)
\ba\label{nonan}
\delta\Sigma\sim x^{2d-4},
\ea
gives us the scaling dimension $d$ of the dual operator ${\cal O}$. 

In terms of the $\Sigma,\chi_p$ (or $\Sigma,\psi_p$) the 5D equations of motion~(\ref{EOM}) decouple:
\ba\label{eqSigma}
x\frac{d}{dx}\Sigma(x)&=&(x-c_M)\Sigma^2(x)+2c_D \Sigma(x)+x+c_M\\\no
\partial_z\log(a^2\tilde\chi_p)&=&(aM_M-p)\Sigma-aM_D.
\ea
Similarly, the class of boundary conditions of the form $\psi=($const$)\chi$ simply reads $\Sigma=$ const. As expected from AdS/CFT, the quantity $\Sigma$ encodes all information about the spectrum of the 4D theory.

The decomposition into local and non-local parts is very useful because it allows us to isolate the differential equation satisfied by $\delta\Sigma$:~\footnote{The terms we neglected are either local or subleading compared to~(\ref{nonan}) as long as $c\propto c_D-c_Ma_0\neq0$. The limit $c=0$ is associated with the regime in which the dual operator has dimension $d=2$, in which logs appear in the expansion in~(\ref{nonan}) rather than powers. For our purposes it will be sufficient to discuss theories with $d\neq2$, and therefore eq.~(\ref{eqnonan}) is accurate. 
}
\ba\label{eqnonan}
x\frac{d}{dx}\Sigma(x)&=&(x-c_M)\delta\Sigma^2+2(c_D-c_Ma_0)\delta\Sigma+O(x\delta\Sigma,x)
\ea
The analytic part satisfies the same equation as in the first line of~(\ref{eqSigma}), and may be solved iteratively. 

Let us first consider the case $d>2$, in which the quadratic part $\delta\Sigma^2$ in~(\ref{eqnonan}) is subleading in a small $x$ expansion. In this case~(\ref{nonan}) and (\ref{eqnonan}) immediately give $d=2+(c_D-c_Ma_0)$. We are hence interested in $a_0=\Sigma_{\rm poly}(0)$, which is easily found to be one of the two solutions of the second order equation $-c_Ma_0^2+2c_Da_0+c_M=0$:
\ba
a_0=\frac{c_D}{c_M}\mp\frac{\sqrt{c_D^2+c_M^2}}{c_M}.
\ea
The different signs characterize two distinct branches,~\footnote{When $c_D\neq0$ one may find it useful to characterize the ``positive" or ``negative" branches by the sign of $c_D$ -- say, by identifying $\pm={\rm sign}(c_D)$ in $a_0$ --, such that a smooth limit $c_M\to0$ exists. However this convention is not appropriate when $c_D=0$, in which case we face the fact that the two branches exist independently of the sign of $c_D$.} and
\ba
c_D-c_Ma_0=\pm c~~~~~~~~~~c\equiv\sqrt{c_D^2+c_M^2}.
\ea
Note that the sign of $c_M$ drops out because unphysical, as emphasized below~(\ref{action}).

We have thus found that in the ``+" branch the scaling dimension $d$ of the boundary operator is given by
\ba\label{+}
d=2+\sqrt{c_D^2+c_M^2}~~~~~~~~~~~~~({\rm~``+"~branch~~\&~~}J=\chi).
\ea
This result generalizes the standard formula obtained when $c_M=0$, $c_D>0$~\cite{Henningson:1998cd}\cite{Mueck:1998iz}\cite{Contino:2004vy}. Note that the fact that $\Sigma(0)\neq0$ implies that the source field $J$ has acquired a mass. This means that we should not expect any UV-localized zero mode in the 4D reduction, except in cases where the mass is finely-tuned away by appropriate boundary conditions.

In the negative branch both quadratic and linear terms in $\delta\Sigma$ must be retained. We therefore find the solution of~(\ref{eqnonan}):
\ba\label{nonansol}
\delta\Sigma_-=\frac{2c(2c-1)}{(1-2c)c_M+2c x+Cx^{+2c}}~~~~~~~~~~~(c\neq\frac{1}{2}),
\ea
with $C=O(1)$ a constant of integration determined by the BC in the IR region (we will comment on the limit $c=1/2$ shortly). We stress that this equation is also valid for the positive branch if we substitute $c\to-c$. Let us here focus on the negative branch and first consider the standard case $|c_M|\ll1$. 

When the Majorana mass is neglected, our solution~(\ref{nonansol}) scales as $\delta\Sigma_-=-2c(1-2c)x^{-2c}/C+\cdots$ for $c<1/2$, implying $d=2-c>3/2$. In the limit $c=1/2$ the correlator becomes $\delta\Sigma_-=1/(C'x-x\log x)+\cdots$, which develops a tachionic pole and clearly does not correspond to the propagator of a free fermion. What happens is that the prescription we are adopting, with $\chi$ identified with the source, breaks down in the negative branch when $c\geq1/2$. A sensible interpretation here might be given by changing prescription, or by introducing a finite UV cutoff as done in~\cite{Contino:2004vy}.

While perhaps a bit less evident, the same correspondence holds for $c_M\neq0$, provided we identify $c_M$ with the coefficient of a mass operator for ${\cal O}$.  Assuming $c<1/2$ the leading non-local term of $\delta\Sigma$ in the negative branch is
\ba\label{-}
\delta\Sigma_-\propto-\frac{2c}{(1-2c)C}\frac{x^{2c}}{c_M^2}+\cdots,
\ea
where again the sign of $c_M$ does not enter. This is precisely the leading non-analytic term in the 2-point function of a CFT operator of dimension $d=2-c>3/2$ deformed by a {\emph{mass term}} $\sim c_M/z_{\rm UV}^{4-2d}$. The mass is a {\emph{relevant}} deformation in a CFT with $d<2$, and cannot be neglected at low energies. (For $d>2$ the mass $\propto c_M$ has no effect and one would recover the ``positive branch" result of eq.~(\ref{+}).) We conclude that 
\ba\label{--}
d=2-\sqrt{c_D^2+c_M^2}~~~~~~~~~~~~~({\rm~``-"~branch~~\&~~}J=\chi). 
\ea
Our interpretation of the negative branch with $c<1/2$, and in particular eq.~(\ref{--}), extends~\cite{Contino:2004vy} to the case $c_M\neq0$.

A discussion of the negative branch with $c>1/2$ may be obtained by changing prescription and identifying $\psi$ with the 4D source. The corresponding dual picture is straightforwardly obtained by observing that in this case the CFT correlator is proportional to $1/\Sigma$, and that the equation~(\ref{eqSigma}) has the following symmetry:
\ba
\Sigma\to-\frac{1}{\Sigma}~~~~~~c_D\to-c_D~~~~~c_M\to-c_M.
\ea
This tells us that the ``$+(-)$" branch with $J=\chi$ corresponds to the ``$-(+)$" branch where the source is taken to be $J=\psi$. 

We end this section noting that when $c_M\neq0$ the source can be a general combination of both $\chi,\psi$. The dual description in this case may be obtained using the results of this appendix.

%%%%%%%%%%%%%%%%%%%%%%%%%%%%%%%%%%%%%

 \end{document}